\documentclass{emulateapj}
\usepackage{apjfonts}

\def\hi{\ifmmode {\rm H}\,{\sc i}~ \else H\,{\sc i}~\fi}

\def\zphot {z_{\rm phot}}
\def\mmsun {M_\star/M_\odot}

\slugcomment{Accepted for publication in ApJ Letters}

\shorttitle{Massive galaxy pair fractions to $z=2$}
\shortauthors{Williams et al.}

\begin{document}

\title{The diminishing importance of major galaxy mergers at higher redshifts}

\author{Rik J. Williams\altaffilmark{1,2}, 
        Ryan F. Quadri\altaffilmark{1,2,3}, 
	Marijn Franx\altaffilmark{2}}
\altaffiltext{1}{Carnegie Observatories, 813 Santa Barbara St., Pasadena,
CA 91101, USA}
\altaffiltext{2}{Leiden Observatory, Leiden University}
\altaffiltext{3}{Hubble Fellow}

\email{williams@obs.carnegiescience.edu}

\begin{abstract}
Using mass-selected
galaxy samples from deep multiwavelength data we investigate the incidence 
of close galaxy pairs between $z=0.4-2$.  Many such close pairs 
will eventually merge, and the pair fraction is therefore 
related to the merger rate.  Over this redshift range the mean pair 
fraction is essentially constant (evolving as 
$f_{\rm pair}\sim (1+z)^{-0.4\pm 0.6}$) with about $6\pm 1$\% of massive galaxies
having a $1:4$ or greater companion within $30h^{-1}$\,kpc.  Assuming 
the timescale over which pairs merge is not a strong function of redshift, 
this implies a similarly constant merger rate (per unit time) out to $z=2$.  
Since about three times as much cosmic time passes at $z<1$ as between 
$z=1-2$, this implies that correspondingly more mergers occur in the 
low-redshift universe.  When minor companions (1:10 mass ratio or greater) 
are included, the pair fraction increases to $\sim 20$\% and still does 
not evolve strongly with redshift.  We also use a rest-frame color 
criterion to select pairs containing only quiescent galaxies (major ``dry 
merger'' progenitors), and find them to be similarly rare and constant
with $4-7$\% of massive quiescent galaxies exhibiting a nearby companion.
Thus, even though other studies
find major mergers to be relatively uncommon since $z=1$, our 
results suggest that few additional mergers occur in the $1<z<2$ range and 
other mechanisms may be required to explain the mass and size growth of 
galaxies over this epoch.
\end{abstract}



\section{Introduction}
Most stellar mass in the universe was formed during a relatively
brief chapter in its history, with the cosmic star-formation
rate peaking around $z\sim 2$
\citep[][and references therein]{lilly96,madau96,hopkins06}.  Cosmological 
models and simulations hold that local massive galaxies were 
hierarchically assembled, with massive $z=0$ galaxies built from successive
mergers of lower-mass constituents.  Therefore, although many 
galaxies continue to add mass through star formation at lower redshifts, 
\emph{assembly} processes (such as major mergers and satellite
accretion) should play an increasingly important role at late times
\citep{guo09}.
But recent observations reveal that the true picture is somewhat more 
nuanced: massive ($M_\star\ga 10^{11} M_\odot$) galaxies,
many of which have ceased significant star formation, exist at least up to
$z\sim 2.5$ \citep{kriek06,kriek08b,cassata08,stutz08,williams09,brammer09,brammer11}.  
Although massive galaxies should continue growing through 
mergers, a substantial fraction of their mass was in place at $z=2$.

Such hierarchical processes serve not only to build up the stellar masses of
galaxies, but also affect their structural properties
(e.g.~effective radius, surface density, and mass profile) 
and morphologies.
It now appears that most quiescent galaxies are extremely
compact and dense for their mass at $z\sim 2.3$ \citep[e.g.][]{vdokkum08}, but 
comparably dense objects
are practically nonexistent in the local universe \citep{taylor10}.  
Moreover, the average size
of quiescent galaxies increases (and surface density within the
effective radius decreases) smoothly with time
\citep[e.g.][]{franx08,damjanov09,williams10}.
Gas-poor ``dry mergers'' have been proposed as a mechanism behind this growth,
but the necessary evolution in the mass-size plane (much faster in size
than mass) is difficult 
to achieve in equal-mass merging scenarios \citep{boylan06}.  
Several processes working in concert,
including major mergers \citep[e.g.][]{bell06,shankar11}, the quenching of progressively 
more extended star-forming galaxies at lower redshifts \citep{vdwel09}, and the 
accretion of stellar mass in the outskirts of these compact ``cores'' 
\citep[due to minor mergers;][]{bezanson09,hopkins09} may therefore drive this size evolution.  

Mergers might thus play an important role in the buildup
of massive galaxies, but observational determinations of their
importance are still uncertain.  At high redshifts mergers
can be difficult to detect, though clues may
nonetheless appear in galaxies' morphologies \citep{conselice03} and
resolved velocity distributions \citep{forster09}.
Another promising technique is the use of close galaxy pairs as a proxy
for mergers since these are easily detectable to high redshift 
and, on average, are expected to merge within a relatively short timescale 
\citep[which can be calibrated using simulations;][KW08]{bundy09,lin11,kitzbichler08}.  

Here we employ a deep 
$0.7$\,deg$^2$ multiwavelength survey to determine pair fractions
between $z=0.4-2.0$.
First we present a brief summary of the data 
used in this sample, then the method for deriving galaxy pair fractions
and the conversion to merger rates, and finally we discuss the implications
of the pair fraction evolution for galaxy formation.
AB magnitudes and cosmological parameters $h=0.7$, $\Omega_M=0.3$, and
$\Omega_\Lambda=0.7$ are used throughout.

\section{Data and Derived Quantities}
For this study we use a catalog (Williams et al.~2012, in preparation)
compiled from the UKIDSS Ultra-Deep
Survey \citep[UDS;][]{lawrence07,warren07} 
Data Release 8 and supplementary data.  The datasets and techniques 
employed for generating this catalog are similar to those we used
with the UDS DR1 in \citet{williams09}; a brief summary 
follows.  \emph{Source Extractor} v2.5.0 \citep{bertin96} was run 
in dual-image mode to detect sources in the UDS DR8 K-band mosaic,
measuring fluxes from a series of deep optical/NIR images: $u^\prime$ from 
archival CFHT data, $BVRi^\prime z^\prime$ from 
the Subaru-XMM Deep Survey \citep[SXDS;][]{furusawa08}, and $JHK$ from 
the UDS DR8, all convolved to the same point-spread function.  Because of
\emph{Spitzer}'s much larger point-response function, 
we followed the deblending technique described by \citet{labbe06} and
\citet{wuyts07} to extract matched $3.6\mu$m and $4.5\mu$m fluxes from 
deep \emph{IRAC} imaging in the UDS field (SpUDS; 
PI J.~Dunlop).  Objects falling near bad
pixels in the optical or near-IR images were excluded, as were those 
with no optical coverage, resulting in an effective image area of 
$\sim 0.7 {\rm deg}^2$.

From this updated catalog we calculate photometric redshifts with EAZY 
\citep{brammer08},
interpolate rest-frame colors with InterRest \citep{taylor09}, 
and derive galaxy masses with FAST \citep{kriek09}.  
The photometric
redshifts agree to $\sim 2$\% with spectroscopic redshifts
in the field; most catastrophic outliers are efficiently removed
via a $\chi^2$ cut.
In calculating the masses we assumed a \citet{salpeter} initial mass 
function (IMF), solar metallicity,
and \citet{bc03} stellar population models, and
applied a factor of $-0.2$\,dex to the masses to bring them in line
with a \citet{kroupa01} IMF \citep[as in][]{franx08,williams10}.  
Although uncertainties on photometrically-derived quantities can be 
substantial, in IR-selected galaxy samples stellar masses and mass-to-light
ratios are relatively robust \citep{kriek08a,muzzin09}.  The catalog has a
95\% point-source completeness limit of $K<24.5$; following
\citet{marchesini09}, the corresponding \emph{mass completeness} is estimated by
scaling each galaxy to the mass it would have at a brighter 
$K=24.0$ limit (to account for the fact that galaxies are not point 
sources); the 75th percentile of scaled galaxy masses thus provides
an estimate of the 75\% mass completeness limit, which we 
estimate as $\log M_\star=9.8$ at $z\sim 2$ for red galaxies 
($U-V_{\rm rest}>1.5$).

\section{Pair fraction measurements}
\subsection{Major ``wet'' and ``dry'' merger candidates} \label{sec_pairs}
We searched the catalog for pairs of galaxies with projected transverse 
separations $<30 h^{-1}$ proper kpc, mass ratios
of $1:4$ or greater, and require that the photometric redshifts
of the galaxies in each pair lie within 
$\left|z_1-z_2\right|/(1+z_1)<0.2$ (where the subscripts 1 and 2 refer
to the more and less massive galaxy in each pair, hereafter denoted 
``primary'' and ``secondary'' respectively).  This large redshift separation
ensures that few physically-associated pairs are missed \citep[according
to the analysis described in][]{quadri10}.
From inserting simulated pairs of point sources into the detection image, we 
find that SExtractor successfully
deblends pairs with separations $\gtrsim 1\farcs 2$; we thus
exclude those with $<13$\,kpc separations (equivalent to 
1\farcs 5 at $z\sim 2$, to account for extended galaxy profiles).
Only pairs containing one primary galaxy 
above $M_1>3.2\times 10^{10} M_\odot$ and a secondary 
within the $1:4$ mass ratio constraint, i.e. $0.25\le M_2/M_1<1$, are 
counted.  The 
constraints on $M_1$ and $M_2/M_1$ ensure that 
secondary galaxies are always above our $\log M_2>9.8$ completeness limit 
for red galaxies at $z\sim 2$.  Finally, the constituent galaxies within 
pairs are classified as 
either star-forming or quiescent based on the rest-frame color criteria of
\citet{williams09}; each pair thus represents a candidate 
``dry merger'' (i.e.~quiescent-quiescent galaxy pair) or 
``wet merger'' (containing at least one star-forming galaxy).

Even with the relatively small angular separations and the 
requirement that galaxies within a pair be close in redshift, some
apparent pairs are due to chance alignments. 
We correct for this by randomizing the positions of all galaxies 
over the survey area (while
retaining the galaxies' masses, redshifts, and quiescent/star-forming 
classifications), and re-running
the pair-finding algorithm on this randomized catalog with the same 
selection parameters \citep[see also][]{quadri10}.  This ``mock catalog''
process is repeated 20 times to reduce the uncertainty in the
contamination rate, and the mean 
number of random pairs is subtracted from the total pair 
counts for each primary and secondary galaxy type.

\begin{figure}
\plotone{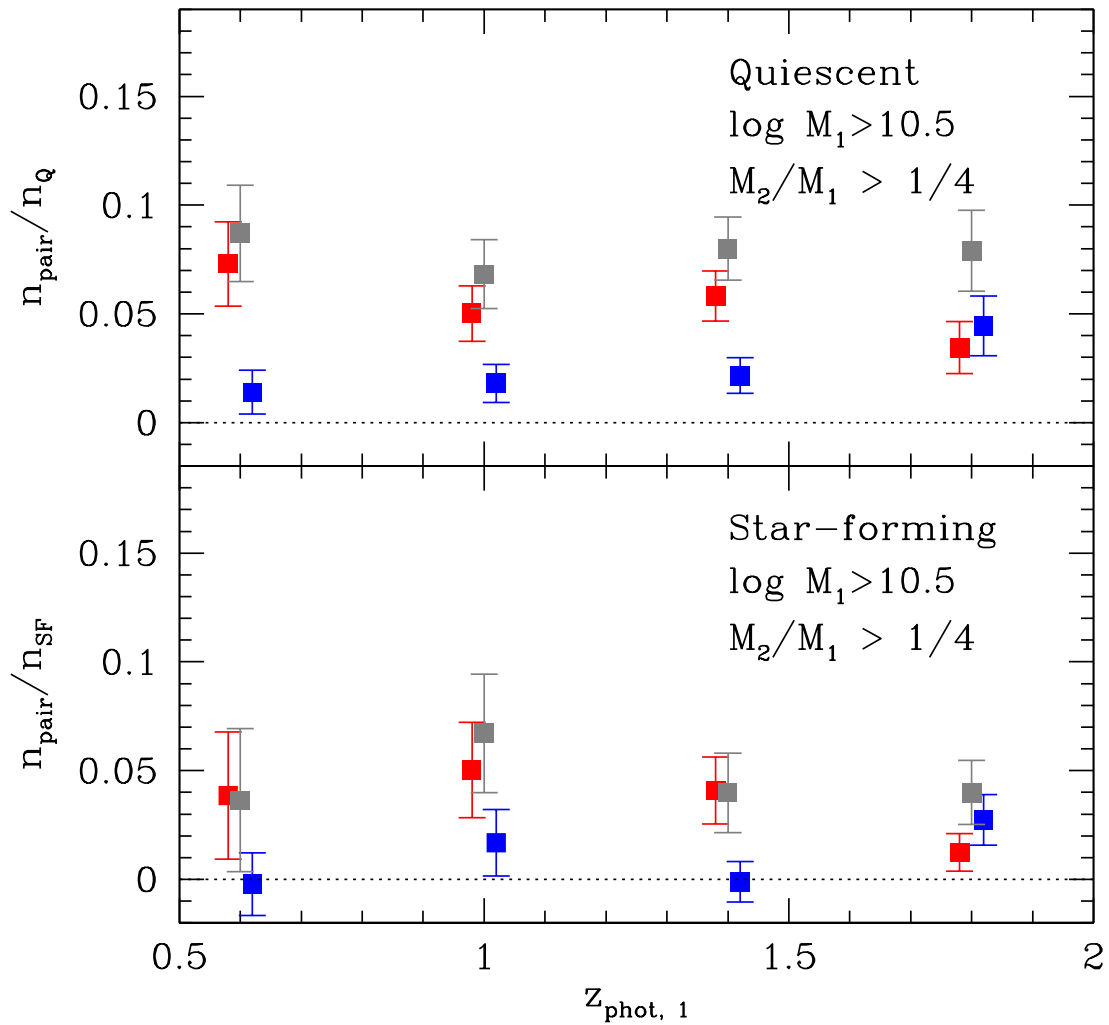}
\caption{Evolution of the massive galaxy pair fraction, defined as the
fraction of massive ``primary'' galaxies ($\log(M_1/M_\odot)>10.5$) which
have close companions within a 1:4 mass ratio (i.e.~$0.25\le M_2/M_1 < 1$;
the average mass ratio is $M_2/M_1\sim 0.4$).
The top and bottom panels show this fraction for quiescent 
and star-forming primary galaxies respectively.
Red, blue, and grey points denote
the fractions of quiescent, star-forming, and total \emph{secondary} galaxies,
respectively; thus, the red points in the top panel show the prevalence
of major ``dry merger''  progenitors.  The total pair fraction is remarkably
constant over this redshift range.
\label{fig_pairs} }
\end{figure}

These corrected pair fraction measurements are listed in
Table~\ref{tab_pairs} and shown in Figure~\ref{fig_pairs}.  
We define the pair
fraction $f_{\rm Xy}=N_{\rm Xy}/N_{\rm X}$ as the fraction of ``primary''
galaxies of type X
which have a less-massive ``secondary'' companion of type y; e.g., $f_{\rm Sq}$
is the fraction of $\log (M_\star/M_\odot)>10.5$ star-forming galaxies
with quiescent galaxy companions (whose masses are $>0.25 M_{\rm primary}$),
counting each pair only once even if
both members are above the primary mass threshold.  The total pair
fraction regardless of galaxy type, $f_{\rm all,all}$, is also given.
Assuming these pairs 
merge at some point in the future, this fraction is related to 
the number of merger \emph{descendants} -- i.e. how many mergers a 
massive $z=0$ galaxy is likely to have undergone -- as a function
of redshift.

\subsection{Mass dependence and minor pairs}
\begin{figure}
\plotone{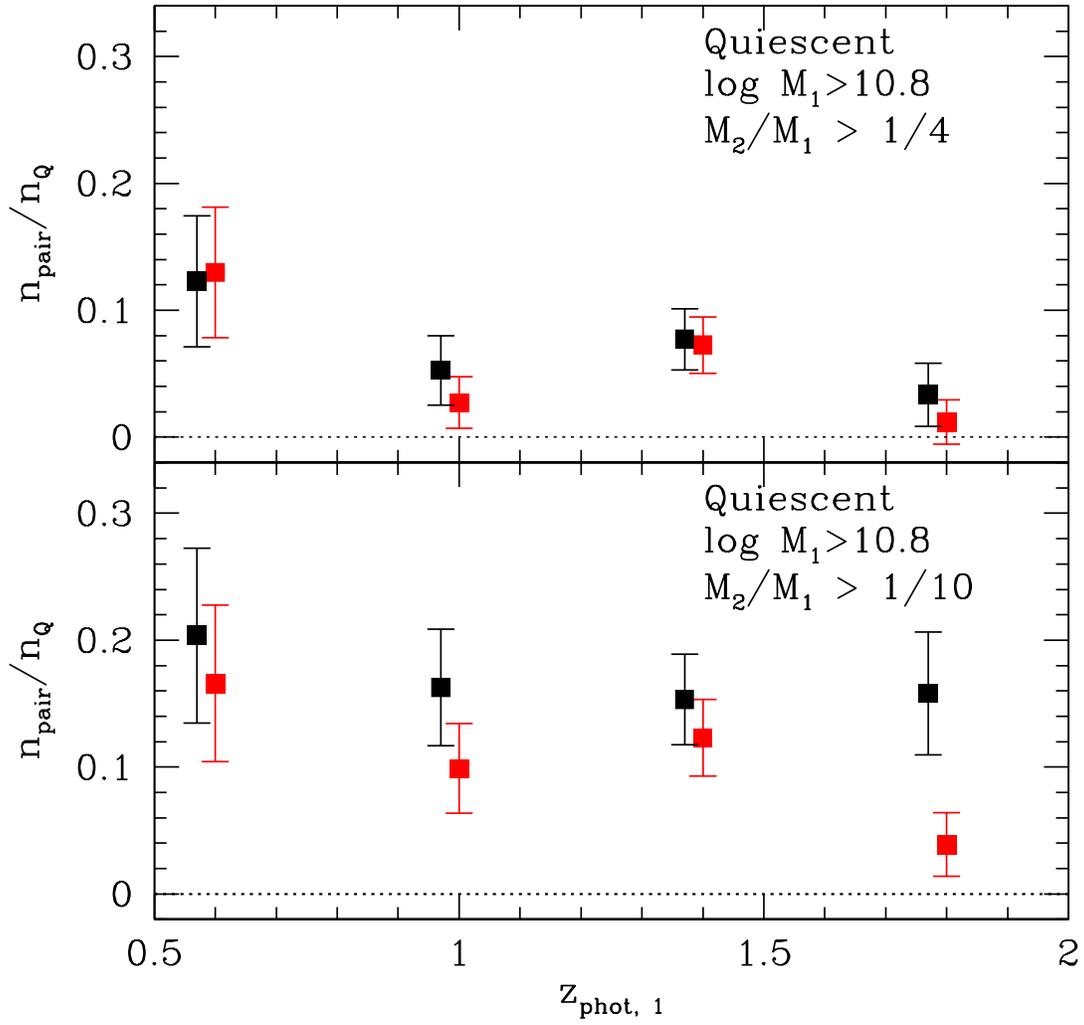}
\caption{Comparison of pair fractions with two different mass ratios
($M_2/M_1>1/4$ and $M_2/M_1>1/10$) as a function of
redshift.  Here all ``primary'' galaxies are quiescent and 0.3 dex higher in mass
than in Figure~\ref{fig_pairs}; black points
show the fraction of primary galaxies with any close companions within
the given mass ratio, while red points denote quiescent secondary galaxies
(``dry mergers'') only.  In both mass ranges the pair fraction is 
constant or weakly declining with redshift.
\label{fig_minpairs}}
\end{figure}

While the pair sample in Table~\ref{tab_pairs} makes
full use of the galaxy catalog's dynamic range, it is also instructive
to consider higher-mass primary galaxies: both to determine whether
the pair fraction depends on mass \citep[e.g.][]{bundy09,bernardi11},
and to investigate lower-mass companions which may
represent minor mergers.   The procedure described above is therefore
repeated for primary galaxies with $\log(\mmsun)>10.8$ and companions
within 1:4 and 1:10 mass ratios;  these fractions are shown
in Figure~\ref{fig_minpairs}.  Since star-forming galaxies are
rare at this mass, only pairs with quiescent primary galaxies are
included in this plot.
Over all redshifts the 1:4 pair fractions are similar to those at
$\log(\mmsun)>10.5$ shown in Figure~\ref{fig_pairs}; this is not
surprising, since these samples are not widely separated in mass and
the uncertainties are large.
Unfortunately, the dynamic range of the catalog does not allow
a detailed analysis of mass dependence due to the relative rarity of very
massive galaxies and the need to maintain a ``secondary'' sample above 
the survey completeness limit.

When minor companions are included, the pair fraction increases 
substantially: while only a few percent of these massive galaxies have a
close neighboring galaxy of comparable mass, 15-20\% have 
companions within a $1:10$ ratio (Figure~\ref{fig_minpairs}, bottom panel).  Since high-mass 
galaxies are more likely to be quiescent, 
this increased pair fraction is largely due to star-forming companions;
however, with the inclusion of lower-mass secondary galaxies
there even appears to be a marginal increase in the ``dry pair''
fraction at each redshift.  Although the ``minor pair fraction'' is
substantially enhanced, this does not necessarily suggest
a high minor merger rate since satellite mass and dynamical friction
timescale are expected to be inversely correlated (KW08).

\subsection{Inferred merger timescales and rates}
Galaxy pairs only represent merger {\it candidates}, as some 
apparent pairs will not coalesce even over many Gyr while others 
rapidly merge, depending on orbital properties, projection effects,
and other internal factors.  Thus, converting from pair fractions
to merger rates requires an average timescale,
\begin{equation}
t_{\rm merg} = N_{\rm pair}/\dot{N}_{\rm merger}
\end{equation}
where $N_{\rm pair}$ is the number of close pairs within a given redshift
bin and $\dot{N}_{\rm merger}$ is the number of mergers per unit
time within the same galaxy population.  
One theoretical parametrization of this timescale was derived from the 
Millennium Simulation by KW08 as a function of galaxy mass, redshift,
and maximum projected separation.  Specifically,
their formula for pairs with projected separations 
$\Delta r< 30 h^{-1}$\,kpc, $M_2/M_1>1/4$,
and photometric redshifts gives a mean merger timescale of
\begin{eqnarray}
\langle t_{\rm merg}(h^{-1} {\rm Myr})\rangle ^{-1/2} = \nonumber 0.0189-
9.47\times 10^{-4} z \\
+ 6.71\times 10^{-3} \left[\log (M_\star /h^{-1} M_\odot)-10\right]
\end{eqnarray}

\begin{figure}
\plottwo{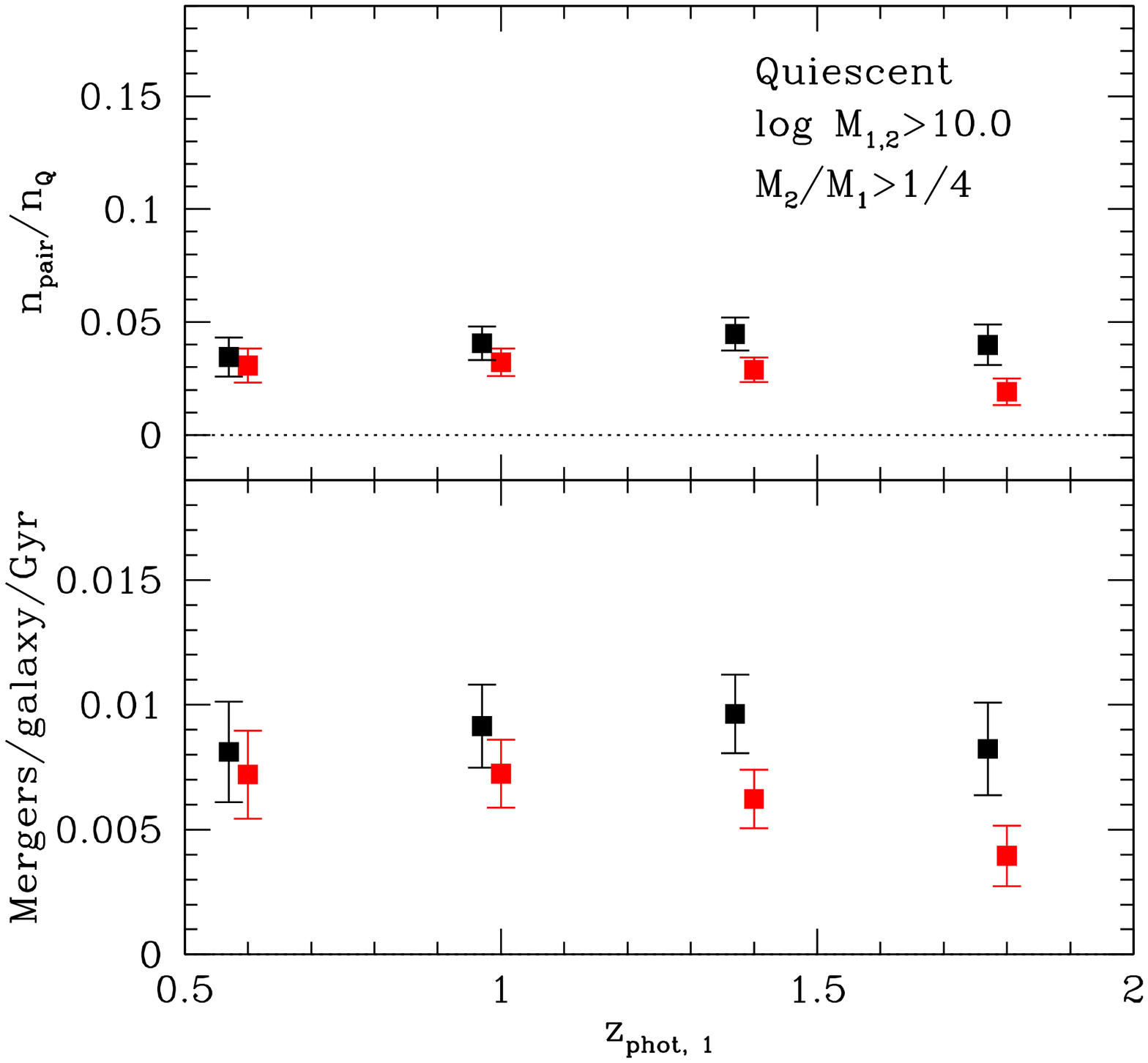}{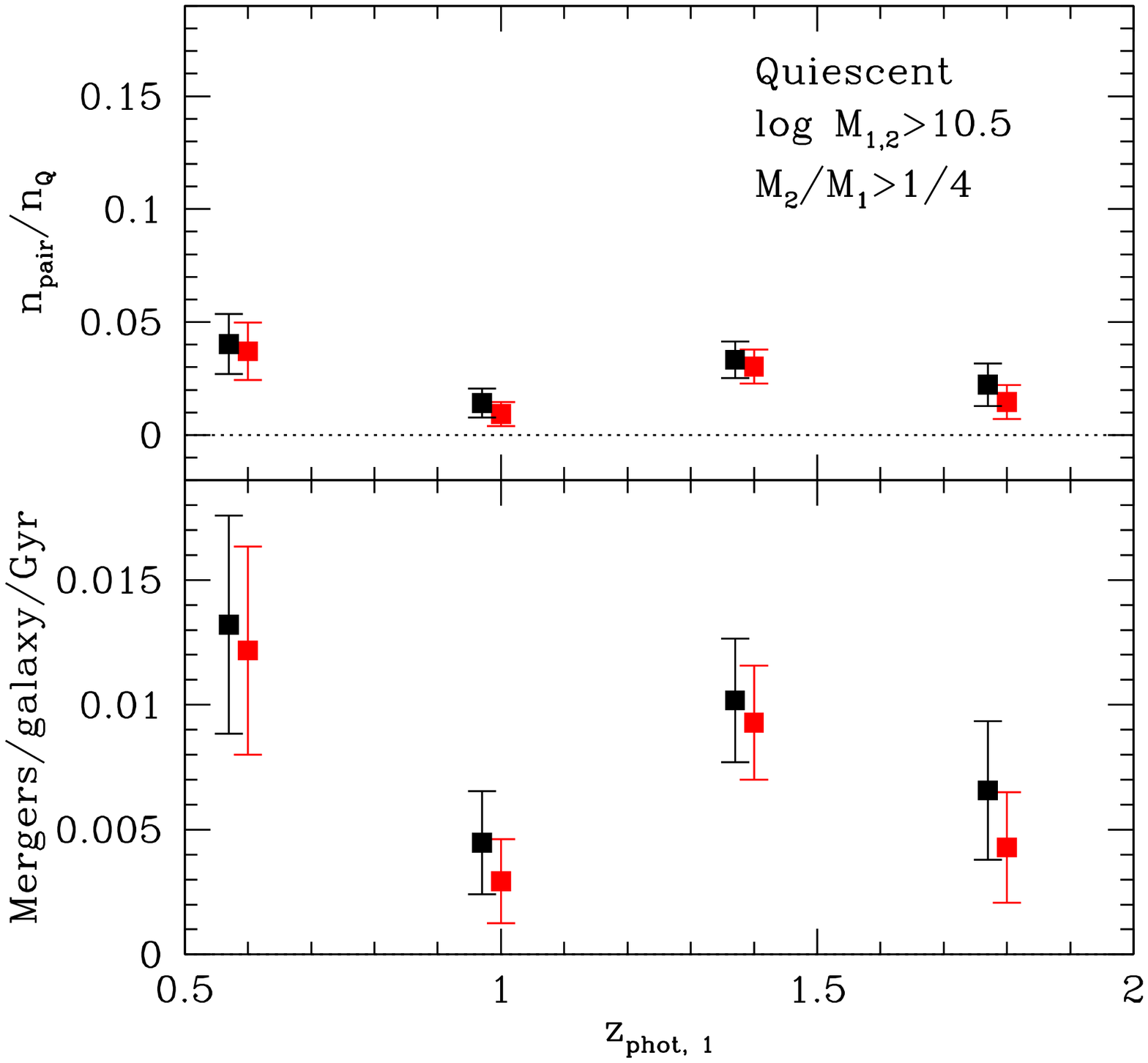}
\caption{Top panels: Pair fractions of quiescent primary galaxies
determined using the same mass threshold 
for both primary and secondary galaxies.  This is the selection method
used in the theoretical merger timescale calibration 
developed by KW08.  Bottom panels: Merger rates
computed from the measured pair fractions
and KW08 timescales.  In all panels, red points denote quiescent-quiescent
galaxy pairs (``dry mergers'') while black points include both
star-forming and quiescent secondary galaxies.  The left and right 
panels show the same calculation for mass limits of
$\log M\star/M_\odot=10.0$ and 10.5 respectively.  Note the fractions in the
right panel differ from Figure~\ref{fig_pairs} due to the higher
mass limit imposed on the secondary galaxies; in both cases the inferred
merger rates are essentially constant and quite low.
\label{fig_mergers}}
\end{figure}

In the KW08 formalism the same mass limit is imposed on both 
primary and secondary galaxies, while our sample includes lower-mass
companions down to the survey limit.
The resulting KW08 pair mass ratios
depend on primary galaxy mass (i.e., closer to equal-mass near the 
survey limit).  For consistency, we thus repeat the above exercise 
with their selection method.
Two mass thresholds are applied:
$\log (M_\star/M_\odot)>10.5$ as before, and $\log (M_\star/M_\odot)>10.0$
to take full advantage of the survey.
These fractions, shown in the top panels of Figure~\ref{fig_mergers}, are 
somewhat lower than in Figure~\ref{fig_pairs} where secondary 
galaxies below the primary mass threshold are included.  

Figure~\ref{fig_mergers} (bottom) shows the specific merger
rates, calculated with eq.~(2), from $z=0.4-2$ at the two mass 
thresholds.  Adopting the KW08 timescale, 
mergers are relatively rare: only about $0.5-1$\% of massive quiescent galaxies
merge with quiescent companions each Gyr, and fewer than 10\% of galaxies
over this redshift range will undergo major mergers.  This is
due both to the rarity of massive galaxy pairs and to the long effective
merger timescales from KW08, about 2-3.5 Gyr
depending on mass and redshift.  Although the KW08 formalism is most
applicable to our specific pair selection parameters, a variety of
timescales have been employed in the literature; this
is discussed further in \S\ref{sec_timescale}.

\section{Discussion}
\subsection{Connecting pairs to mergers} \label{sec_timescale}
Close pairs of galaxies are easily identified and detectable to
high redshift, and are therefore in principle a robust way of identifying
systems that may merge within a relatively short timescale.  But it is less 
straightforward to determine 
this timescale and convert the measured pair fractions to merger
rates.

Several estimates of the merger timescale have been used
in previous work.  \citet{bell06} assume that galaxy pairs merge within
roughly one orbital time, in their case $\sim 0.4$\,Gyr,
while the KW08 estimate is nearly an order of magnitude larger for the
galaxy masses considered here.  These
fundamentally change the interpretation of the measured pair fractions:
with the \citet{bell06} timescale, major mergers play a significant role
in the assembly of massive galaxies over time; assuming KW08, only
about 20\% of massive quiescent galaxies have undergone major mergers since
$z=2$, with $\sim 2/3$ of this occurring at $z<1$.  This disagreement is 
largely a result of the KW08 analysis 
including physically-associated galaxy pairs which are at 
relatively large real distances despite having close projected separations, 
and therefore merge only after a long period (or not at all).  
Nonetheless, if the merger timescale isn't a strong function of
redshift, the unchanging pair fraction we measure reflects a similarly 
constant merger rate since $z=2$.

\subsection{Comparison to previous work}
The pair fractions derived here are in broad agreement with
previous work at $z\la 1$.  \citet{bundy09} find a low ($\sim 4$\%), non-evolving 
fraction of massive galaxies in ``major pairs'' at $z<1.2$, with
$10^{11}$\,M$_\odot$ galaxies more likely to have close companions
than those with $10^{10}$\,M$_\odot$.  Given their higher
mass limit and smaller search radius ($20 h^{-1}$\,kpc), our $z<1$ measurements
appear to be in agreement: a total pair fraction of $\sim 4-8$\%,
depending on whether the primary galaxies are star-forming or quiescent,
and no strong evolution in the fraction.  In addition, we confirm their
reported higher incidence of ``dry pairs'' at lower redshifts, simply
due to the coincident increase in the number density of massive quiescent
galaxies.  Between $0.4<z<0.8$,
 \citet{bell06} report a pair fraction of $5\pm 1$\% for galaxies
above $2.5\times 10^{10}$\,M$_\odot$, also in agreement
with our pair fraction measurement at the same redshift.

One common theme in these studies is the rarity of dry mergers:
even \citet{bell06}, with their short assumed timescale,
find that only $\sim 50$\% of massive galaxies have undergone major mergers
since $z=0.8$; \citet{bundy09} estimate 30\% at the high-mass
($10^{11}$\,M$_\odot$) end \citep[but see][]{padilla11}.  In their analysis of the environmental
dependence of merger rate, \citet{lin11} estimate a somewhat higher
rate in \emph{high-density environments}: $1.2\pm 0.3$ major dry mergers per
galaxy since $z=1$, perhaps not surprising since such environments
harbor a larger fraction of quiescent galaxies \citep[e.g.][]{kauffmann04,quadri11}.
However, these only account for $38\pm 10$\% of
the mass accretion of these galaxies.
If the merging timescale doesn't vary strongly with redshift,
our results suggest that the major merger rate at $1<z<2$ is comparable to 
that at $z<1$; since twice as much cosmic time passes in the latter epoch
than the former, this in turn implies that most major mergers
occur below $z\sim 1$.

\subsection{Major vs.~minor mergers}
Even when short timescales are assumed, dry mergers from $z=2$ to the present
occur perhaps once or twice per galaxy at most.
Since strong size and mass growth are
nonetheless seen over this same redshift interval 
\citep[e.g.][]{franx08,williams10}, 
it appears that major mergers are \emph{not}
the primary driver behind the observed evolution.  
Indeed, if major dry mergers were the primary driver behind the smooth
evolution in galaxy sizes and surface densities seen over $z=0-2$, 
a much larger number would be required 
to eliminate the compact quiescent galaxy population
by $z=0$ \citep{taylor10}.  
Even if mergers are more common than the KW08 timescale implies, 
they still may not account for the evolution of the \emph{mass-size} relation 
since major mergers increase galaxies' masses and sizes at similar rates.

Another explanation is that galaxies undergo many minor mergers or accrete
low-mass satellites.  \citet{naab09} find that minor mergers
are more efficient (per unit secondary galaxy mass) at increasing galaxy radii than
major mergers.  This accretion 
scenario is also attractive because it depends only 
weakly (if at all) on whether or not the central galaxy is forming stars, 
and might also partially explain the size growth observed in massive 
star-forming galaxies \citep{williams10}.
As shown in Figure~\ref{fig_minpairs}, the inclusion of minor companions
increases the pair fraction to $15-20$\% \citep[see also][]{lopez11}.  However, the merging 
timescale of these minor galaxies is
not well-understood: on one hand their dynamical friction timescales
are longer than more massive satellites, but tidal
effects may hasten the incorporation of minor satellites into a massive
galaxy's extended halo.  

\section{Conclusions}
From deep mass-selected samples, we have performed the first
analysis of major and minor galaxy pairs out to $z=2$, 
distinguishing between ``dry'' and ``wet'' merger candidates.  
``Dry'' pairs are relatively rare: only $\sim 3-7$\% of 
massive quiescent galaxies have a quiescent companion of comparable mass
(with an average companion mass 40\% that of the primary). 
This fraction increases slightly at lower redshifts, but the
total pair fraction (including both star-forming and quiescent
galaxies) remains essentially constant, evolving as 
$f_{\rm all,all}\sim (1+z)^{-0.4\pm 0.6}$.  Minor companions
are significantly more common, increasing to 15-20\% for $>1:10$ mass
ratios.  However, significant uncertainties remain when converting
pair fractions to merger rates.
Nonetheless, while 
galaxies are often assumed to assemble rapidly at high z, the 
constant pair fraction
suggests that, in fact, massive galaxies undergo more mergers
at $z<1$ than $z=1-2$.  Unless the effective merger timescale is much
shorter than the values assumed here, such mergers represent a rare and 
stochastic process that is 
unlikely to make more than a marginal contribution to the smooth mass and 
size growth seen since $z=2$.  

\acknowledgments
We thank the referee for helpful and constructive comments.
R.J.W. acknowledges support from NSF grant AST-0707417.  Much of this
work was performed during working visits to Leiden, funded by the
Netherlands Organization for Scientific Research (NWO) and the Leids
Kerkoven-Bosscha Fonds.

\begin{deluxetable}{lccccccc}
\tablecolumns{8}
\tablewidth{450pt}
\tablecaption{Pair fractions in the UKIDSS-UDS \label{tab_pairs}}
\tablehead{
\colhead{$\zphot$} &
\colhead{$N_{\rm Q}$} &
\colhead{$N_{\rm S}$} &
\colhead{$f_{\rm Qq}$} &
\colhead{$f_{\rm Qs}$} &
\colhead{$f_{\rm Sq}$} &
\colhead{$f_{\rm Ss}$} &
\colhead{$f_{\rm all,all}$}
}
\startdata

$0.4<z<0.8$ &270 &77  &$0.066\pm 0.017$ &$0.013\pm 0.009$ &$0.035\pm 0.026$ &$-0.001\pm 0.013$ &$0.069\pm 0.016$\\
$0.8<z<1.2$ &495 &163 &$0.046\pm 0.011$ &$0.016\pm 0.008$ &$0.046\pm 0.019$ &$0.015\pm 0.014$ &$0.062 \pm 0.012$\\
$1.2<z<1.6$ &692 &292 &$0.053\pm 0.010$ &$0.020\pm 0.007$ &$0.037\pm 0.014$ &$-0.001\pm 0.008$ &$0.062 \pm 0.010$\\
$1.6<z<2.0$ &376 &387 &$0.031\pm 0.011$ &$0.040\pm 0.012$ &$0.011\pm 0.008$ &$0.025\pm 0.010$ &$0.054 \pm 0.010$\\
\hline

\enddata
\tablecomments{ Uppercase subscripts denote the ``primary'' galaxy type:
quiescent (Q) or star-forming (S);  lowercase 
denote the type of the secondary galaxy within a pair. 
Hence ``$f_{\rm Qs}$'' is
the fraction of massive quiescent galaxies with a star-forming
companion.  Here all primary galaxies have masses above $\log M_1>10.5$, 
secondaries are above $\log M_2>9.9$, and a mass ratio constraint
of $0.25<M_2/M_1<1$ has been imposed.  
}
\end{deluxetable}

\end{document}